\documentclass[twocolumn,aps,prl,superscriptaddress]{revtex4-1}

\usepackage{amssymb,amsmath}
\usepackage{graphicx}

\usepackage{braket}
\usepackage[colorlinks=true,citecolor=blue,linkcolor=blue,linktocpage=true,pagebackref=false]{hyperref}
\hypersetup{colorlinks=true,citecolor=blue,linkcolor=blue,filecolor=blue,urlcolor=blue}
\usepackage{color,soul}

\providecommand{\e}{\text{e}}

\providecommand{\Tr}{\text{Tr}} 
\providecommand{\tr}{\text{tr}}

\begin{document}
\title{Landauer-B\"uttiker approach to strongly coupled quantum thermodynamics: inside-outside duality of entropy evolution}

\author{Anton Bruch}
\affiliation{\mbox{Dahlem Center for Complex Quantum Systems and Fachbereich Physik, Freie Universit\"at Berlin, 14195 Berlin, Germany}}
\author{Caio Lewenkopf}
\affiliation{Instituto de F\'{i}sica, Universidade Federal Fluminense, 24210-346 Niter\'{o}i, Brazil}
\author{Felix von Oppen}
\affiliation{\mbox{Dahlem Center for Complex Quantum Systems and Fachbereich Physik, Freie Universit\"at Berlin, 14195 Berlin, Germany}}

\date{\today}
\begin{abstract}
We develop a Landauer-B\"uttiker theory of entropy evolution in time-dependent
strongly coupled electron systems. This formalism naturally avoids
the problem of system-bath distinction caused by the strong hybridization
of central system and surrounding reservoirs. In an adiabatic expansion
up to first order beyond the quasistatic limit, it provides a clear
understanding of the connection between heat and entropy currents
generated by time-dependent potentials and shows their connection
to the occurring dissipation. Combined with the work required to change
the potential, the developed formalism provides a full thermodynamic
description from an outside perspective, applicable to arbitrary non-interacting
electron systems. 
\end{abstract}

\maketitle


{\it Introduction.$-$}Ongoing progress in nanofabrication raises interest in the thermodynamics
of nanomachines \cite{Millen2016,Pekola2015}, describing the exchange
of heat and work with their environment as well as their efficiencies. The
laws of thermodynamics are extremely successful in characterizing
machines consisting of a macroscopic number of particles by just a
handful of parameters such as temperature and pressure. How these
laws carry over to microscopic systems that consist of few particles
and exhibit quantum behavior, is the central problem of the field
of quantum thermodynamics. At small scales, the thermodynamic variables
necessarily acquire strong fluctuations \cite{Jarzynski2011,Seifert2012a}
and the system-bath distinction becomes fuzzy \cite{Hanggi2008, LilianaDynamical, 
Bruch2016,Ochoa2016}. 
 A crucial quantity in this regard is the entropy which links thermodynamics
and information \cite{Landauer1961,Berut2012,Koski2013,Parrondo2015},
describes irreversibility, and governs the efficiencies of various
energy conversion processes.

Here, we put forward a formalism based on the Landauer-B\"uttiker scattering approach 
to describe the entropy evolution generated by (slow) time-dependent potentials in electronic 
mesoscopic systems strongly coupled to external reservoirs.  
A key advantage of the scattering approach is that it naturally
avoids the system-bath distinction, a ubiquitous problem for theoretical
treatments of the strong coupling regime \cite{Hanggi2008,Campisi2009,
Hilt2009a,Esposito2010a,Strasberg2016,Perarnau-Llobet2016}.

In an elementary thermodynamic transformation, an external agent performs
work on a system by changing its Hamiltonian, constituting a single
``stroke'' of a quantum engine. For electronic nanomachines, this
is achieved by changing the potential in a finite region which is
coupled to electronic reservoirs. This type of machine can for instance be realized
by a quantum dot connected to leads and subject to a time-dependent
gate potential. If the gate potential is changed slowly, the coupling
to the reservoir ensures thermal equilibrium at all times and the
transformation occurs quasistatically. The change of the von-Neumann
entropy
\begin{equation}
{\bf S}[\rho]=-\Tr\left(\rho\ln\rho\right)\,,\label{eq:vonNeumannS}
\end{equation}
associated with the equilibrium state of the system is proportional
to the heat $dQ=Td{\bf S}$ released into the reservoir at temperature
$T$. 

This should be contrasted with the entropy evolution of a closed quantum
system. Its purely unitary time evolution implies that the von-Neumann
entropy remains unchanged at all times. Here we want to discuss the
entropy evolution of simple electronic nanomachines, which combine
fully coherent quantum dynamics with contact to baths and can involve
strong coupling between system and reservoir. In this problem, quantum
effects such as coherences, hybridization, and entanglement are expected
to become important.  
Such electronic nanomachines can be described by the Landauer-B\"uttiker formalism, that
has been very successfully used to understand the conductance \cite{Buttiker1992},  electron 
pumping \cite{Brouwer1998}, heat transport and current noise \cite{Buttiker1992,Moskalets2002,Moskalets2004,AvronPRL01},
entanglement creation \cite{Beenakker2003,Samuelsson2004}, and adiabatic reaction forces \cite{BodePRL,Bode2012,Thomas2012,Thomas2015}  in a variety of mesoscopic
systems.

\begin{figure}[h!]
\includegraphics[width=0.4\textwidth]{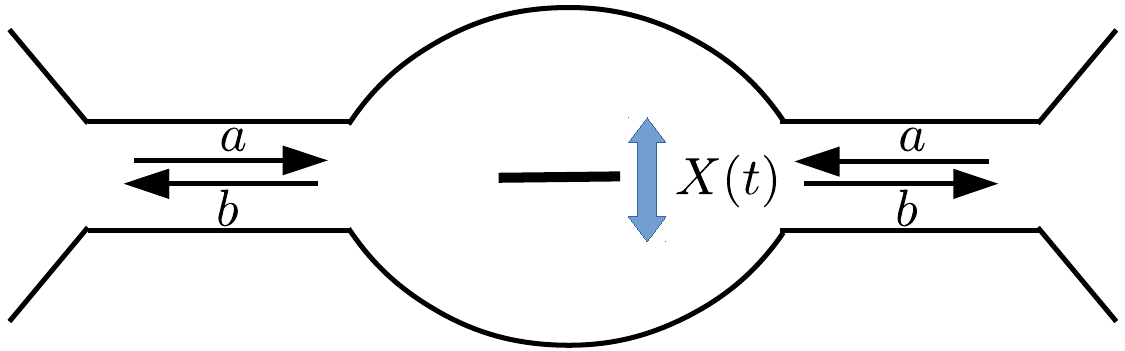}
\caption{The scattering potential in the central region, e.g. a quantum dot,
slowly changed by an external parameter $X(t)$, driving  a net heat and 
entropy current into the leads.
\label{Sketch1}}
\end{figure}

One considers a scattering region connected
to ideal leads, as depicted in Fig.\ \ref{Sketch1}, where non-interacting
electrons propagate freely and under fully coherent quantum dynamics.
Relaxation is accounted for by connecting the leads to electronic
reservoirs at well defined temperatures and chemical potentials, which
determine the distribution of incident electrons. This allows one
to calculate energy or particle currents in the
leads by accounting for in- and outgoing electrons. Invoking energy and
particle conservation, these currents permit one to deduce the change
of energy and particle number in the scattering region from an\emph{
outside perspective.} Since the von-Neumann entropy ${\bf S}$ is conserved
under coherent unitary dynamics, the change of entropy in the
scattering region can also be inferred from the entropy currents carried
by the scattered electrons. The subsequent relaxation processes occur in
the bath. In the Landauer-B\"uttiker formalism the reservoirs are macroscopic 
and relax to equilibrium such that excitations 
entering a reservoir never return \cite{Buttiker1992}.

Several studies have investigated the thermodynamic properties 
of mesoscopic electronic 
systems using the simplest model that captures the difficulties of the strong coupling regime
\cite{LilianaDynamical,EspositoQThermo,Bruch2016}: 
A single electronic level strongly coupled to a free electron reservoir and subject to a slowly
varying gate potential.
The thermodynamic transformations of this system were described from
an \emph{inside perspective}, i.e., in terms of the thermodynamic
variables of the single level. To account for the hybridization
in the strong coupling regime, these thermodynamic functions were
treated with the help of the nonequilibrium Green's function formalism.
This approach was recently found to be limited to a small set of systems
\cite{Ochoa2016}. 


{\it Entropy current carried by scattering states.$-$}We consider a time-dependent scattering region connected to one or
multiple ideal leads, in which the electrons propagate in transverse
scattering channels, and leave the electronic spin degree of freedom implicit.
Electrons in incoming and outgoing channels are described by annihilation operators $a$ and $b$,
 related by the scattering matrix ${\cal S}$, 
\begin{equation}
\left(\begin{array}{c}
b_{1}(\epsilon)\\
\vdots\\
b_{N}(\epsilon)
\end{array}\right)=\int\frac{d\epsilon'}{2\pi}{\cal S}(\epsilon,\epsilon')\left(\begin{array}{c}
a_{1}(\epsilon')\\
\vdots\\
a_{N}(\epsilon')
\end{array}\right)\,.\label{eq:scatteringSchematically}
\end{equation}
Here the subscript $\alpha=1, \cdots, N$ labels the channels and leads.
The leads are connected
to electronic reservoirs, which determine the distribution of the
incoming channels to be 
$\braket{a_{\beta}^{\dagger}(\epsilon)a_{\alpha}(\epsilon')}=
\phi_{\alpha\beta}^{\text{in}}(\epsilon)2\pi\delta(\epsilon-\epsilon')$ in terms of a diagonal distribution matrix 
$\phi_{\alpha\beta}^{\text{in}}(\epsilon)=\delta_{\alpha\beta}f_{\alpha}(\epsilon)$, where
$f_{\alpha}(\epsilon)$ is the Fermi distribution with temperature $T$ and 
chemical potential $\mu_\alpha$.

The particle current in channel $\alpha$ through any cross-section of the 
corresponding lead is obtained by accounting for in- and outgoing electrons 
\cite{Buttiker1992},
\begin{equation}
I_{\alpha}^{N}(t)=\int_{-\infty}^{\infty}\frac{d\epsilon}{2\pi}\,\left\{ \phi_{\alpha\alpha}^{\text{out}}(t,\epsilon)-
\phi_{\alpha\alpha}^{\text{in}}(\epsilon)\right\} \,,\label{eq:Particle_Current}
\end{equation}
where the one-dimensional density of states $\varrho_{\alpha}(\epsilon)=\left[2\pi v_{\alpha}(\epsilon)\right]{}^{-1}$
and the group velocity $v_{\alpha}(\epsilon)$ compensate (we set $\hbar=1$). $\phi^{\text{out}}(t,\epsilon)$
is given by the Wigner transform 
\begin{equation}
\phi_{\alpha\beta}^{\text{out}}(t,\epsilon)=\int\frac{d\tilde{\epsilon}}{2\pi}e^{-i\tilde{\varepsilon}t}
\left \langle b_{\beta}^{\dagger}(\epsilon-\tilde{\epsilon}/2)b_{\alpha}(\epsilon+\tilde{\epsilon}/2) \right\rangle \,.
\label{eq:PhiWigner}
\end{equation}

Similarly, the energy current $I_{\alpha}^{E}$ in channel $\alpha$ reads
\begin{equation}
I_{\alpha}^{E}(t)=\int_{-\infty}^{\infty}\frac{d\epsilon}{2\pi}\,\epsilon\left\{ \phi_{\alpha\alpha}^{\text{out}}(t,\epsilon)-\phi_{\alpha\alpha}^{\text{in}}(\epsilon)\right\} \,.
\end{equation}
The heat current $I_{\alpha}^{Q}=I_{\alpha}^{E}-\mu_{\alpha}I_{\alpha}^{N}$
carried by the electrons in the leads is a combination of the particle
current $I_{\alpha}^{N}$ into the corresponding reservoir with chemical
potential $\mu_{\alpha}$ and the energy current $I_{\alpha}^{E}$.
We can express the total heat current in terms of the diagonal
elements of the distribution matrix $\phi^{\rm out}$
\begin{equation}
I_{\text{tot}}^{Q}(t)=\int_{-\infty}^{\infty}\frac{d\epsilon}{2\pi}\,(\epsilon-\mu)\tr_c\left\{ \phi^{\text{out}}(t,\epsilon)-\phi^{\text{in}}(\epsilon)\right\} \,,
\label{eq:totalHeat}
\end{equation}
where the trace runs over channel and lead space. Here, for simplicity we assume the same chemical potential $\mu$ in all reservoirs. 

To obtain the entropy current, we begin by considering the entropy of a single 
incoming channel. For a given energy the channel can be either
occupied or empty, according to $f_{\alpha}(\epsilon)$, and contributes
with
\begin{equation}
\sigma\left[f_{\alpha}(\epsilon)\right]=-f_{\alpha}(\epsilon)\ln\left[f_{\alpha}(\epsilon)\right]-(1-f_{\alpha}(\epsilon))\ln\left[1-f_{\alpha}(\epsilon)\right]\,\label{eq:Sigma}
\end{equation}
to the system entropy.
By analogy with the particle current, Eq.~\eqref{eq:Particle_Current}, we write the incoming entropy 
current as
\begin{equation}
I_{\alpha}^{S\,\text{in}}=\int_{-\infty}^{\infty}\frac{d\epsilon}{2\pi}\,\sigma\left[f_{\alpha}(\epsilon)\right]\,.
\label{eq:EntropyCurrentSingleChannel}
\end{equation}
Hence, as expected \cite{Pendry1983}, each of the incoming spin-resolved channels carries an 
entropy current of $\pi T/6$ towards the scattering region.

Scattering redistributes the electrons between the outgoing channels, thereby
modifying the entropy flow into the leads.
The scattering-induced correlations between outgoing scattering states  \cite{Moskalets2002,Moskalets2004} are encoded 
in the non-diagonal distribution matrix $\phi_{\alpha\beta}^{\text{out}}(t,\epsilon)$
for the outgoing electrons. 
As we show below, the natural extension of Eq.~\eqref{eq:EntropyCurrentSingleChannel} 
reads
\begin{equation}
I^{S\,\text{in}(\text{out})}(t)=\int_{-\infty}^{\infty}\frac{d\epsilon}{2\pi}\,\text{tr}_c\left\{ \sigma[\phi^{\text{in}(\text{out})}(t,\epsilon)]\right\} \,.
\label{eq:GeneralizedEntropy}
\end{equation}

To motivate Eq.\ \eqref{eq:GeneralizedEntropy} we derive the 
non-interacting fermionic density matrix for a given distribution matrix 
$\bar{\phi}_{\alpha\beta}=
\Tr[\rho c_{\beta}^{\dagger}c_{\alpha}]$.
In the scattering setup the incoming operators describe particles
of an equilibrium reservoir and the outgoing operators are linear
functions of the incoming ones, cf. Eq.\ \eqref{eq:scatteringSchematically}.
Hence, all averages can be calculated via Wick's theorem and the single-particle correlations described by $\phi$ fully determine all expectation
values. 

Our derivation exploits the maximum entropy
principle that yields the most general density matrix given certain 
single-particle correlations \cite{Jaynes1957}. (We obtain the same result following the
approach of Ref.\ \cite{Peschel2003}.) The Lagrangian for maximizing the von-Neumann entropy under the constraints
$\Tr\rho=1$ and $\bar{\phi}_{\alpha\beta}= \Tr[\rho c_{\beta}^{\dagger}c_{\alpha}]$
reads
\begin{align}
{\cal L} & =-\Tr\left[\rho\ln\rho\right]+\sum_{\alpha\beta}\lambda_{\alpha\beta}
\left(\Tr\left[\rho c_{\beta}^{\dagger}c_{\alpha}\right]-\bar{\phi}_{\alpha\beta}\right)\nonumber \\
 & -\gamma\left(\Tr\rho-1\right)\,,\label{L}
\end{align}
where $\Tr$ denotes the many-particle trace over all possible occupations
and $\gamma$ as well as the $\lambda$'s are Lagrange multipliers. 
It is convenient to diagonalize the Hermitian matrix $\bar{\phi}$ and introduce a rotated basis, namely 
\begin{equation}
\bar{\phi}  =U\Lambda U^{\dagger}\quad \mbox{and} \quad c_{\alpha}  =\sum_{c}U_{\alpha c}d_{c}\,,
\end{equation}
where $U$ is a unitary matrix and $\Lambda_{\alpha\beta}=\Lambda_{\alpha}\delta_{\alpha\beta}$ 
is diagonal containing the real eigenvalues of $\bar{\phi}$. In the rotated basis the Lagrangian
${\cal L}$  allows us to maximize the von-Neumann entropy with the
given constraints. This yields the density matrix  
\begin{equation}
\rho=\prod_{\alpha}\left(1-\Lambda_{\alpha}\right)\left(\frac{\Lambda_{\alpha}}{1-\Lambda_{\alpha}}\right)^{\hat{n}_{\alpha}}\,,
\end{equation}
where $\hat{n}_{\alpha}$ is the occupation of mode $\alpha$ in the
rotated basis. We calculate the entropy $\boldsymbol{S}$ 
of this density matrix by summing over all possible occupations in the rotated basis,
\begin{align}
\boldsymbol{S} & =\sum_{\alpha}\sigma[\Lambda_{\alpha}]=\text{tr}\left(\sigma[\Lambda]\right)\,,
\end{align}
where the sum over the diagonal elements of $\Lambda$ is included through the single-particle trace $\text{tr}$. 
Finally, rotating back to the original basis, $\Lambda=U^{\dagger}\bar{\phi} U$, we find the entropy in terms of 
the distribution matrix $\bar{\phi}$,
\begin{align}
\boldsymbol{S} & =\text{tr}\left(\sigma[\bar{\phi}]\right)\,.\label{eq:STrPhi}
\end{align}
For a slowly changing scattering potential, we associate the entropy 
with the time-dependent distribution matrix $\phi_{\alpha\beta}(t,\epsilon)$ of the scattering states in Eq.\ \eqref{eq:PhiWigner}, for which the single-particle trace represents an integral
over energy and a trace $\tr_c$ over channel and lead indices.

By combing in- and outgoing entropy currents we write the total
entropy current into the leads as
\begin{equation}
I_{\text{tot}}^{S}(t)=\int_{-\infty}^{\infty}\frac{d\epsilon}{2\pi}\,\tr_c\left\{ \sigma[\phi^{\text{out}}(t,\epsilon)]-\sigma[\phi^{\text{in}}(\epsilon)]\right\} \,.
\label{eq:totalEntropy}
\end{equation}
In the case of a static scatterer between two biased reservoirs at zero 
temperature, the entropy current can be used to quantify the entanglement 
of outgoing electron-hole pairs created in a tunneling event. 
Indeed, we verify that an immediate generalization of Eq.~\eqref{eq:totalEntropy} reproduces the quantum mutual information between outgoing scattering channels on the left and right as obtained in Ref.~\cite{Beenakker2003} (see \footnote{See Supplemental Material} for details).



{\it Entropy current induced by a dynamic scatterer.$-$}The entropy and heat currents generated by a slowly changing
scattering potential $V\left[X(t)\right]$ are obtained by expanding the scattering
matrix and the outgoing distribution matrix about the frozen configuration
in powers of the velocity $\dot{X}$ \cite{BodePRL,Bode2012,Thomas2012,Thomas2015}. Up to first order, the Wigner transform of the scattering matrix can be expressed in terms of the 
frozen scattering matrix $S$ and its first order correction $A$, ${\cal S}(\epsilon,t)=S+\dot{X}A$.
This
expansion is well motivated in the regime where $X(t)$ changes on a characteristic 
time scale much longer than the electronic dwell time in the scattering region. 
Accordingly, we write $\phi^{\rm out}$ as
\begin{equation}
\phi^{\text{out}}\simeq\hat{I}\,f+\phi^{\text{out}(1)}+\phi^{\text{out}(2)}\,,
\label{eq:ExpansionPhi}
\end{equation}
where $\hat{I}$ is a unit matrix in channel and lead space and the superscript
stands for the order in $\dot{X}$. (We omit time and
energy labels for better readability.) Similarly, we expand  $\sigma[\phi^{\text{out}}(\epsilon)]$ 
up to second order about the uncorrelated equilibrium
\begin{align}
\sigma[\phi^{\text{out}}] & \simeq\hat{I}\sigma\left[f\right]+\hat{I}\frac{d\sigma\left[f\right]}{df}\left(\phi^{\text{out}(1)}+\phi^{\text{out}(2)}\right)
\nonumber \\
 & +\frac{1}{2} \hat{I}\,\frac{d^{2}\sigma\left[f\right]}{df^{2}}\left(\phi^{\text{out}(1)}\right)^{2}\,.\label{eq:SigmaExpansion}
\end{align}
Note that the second order contribution proportional to $d^{2}\sigma\left[f\right]/df^{2}=\left(T\partial_{\epsilon}f\right)^{-1}$
is always negative due to the concavity 
of $\sigma$.

By inserting the above expression in Eq.~\eqref{eq:totalEntropy} we obtain
\begin{align}
I_{\text{tot}}^{S}  =\int_{-\infty}^{\infty}\frac{d\epsilon}{2\pi}\,\tr_c\bigg\{&\frac{\epsilon-\mu}{T}
\left(\phi^{\text{out}(1)}+\phi^{\text{out}(2)}\right)
\nonumber \\ & 
+\frac{1}{2T\partial_{\epsilon}f}\left(\phi^{\text{out}(1)}\right)^{2}\,\bigg\}\,,
\label{eq:IStot}
\end{align}
where we have used that $\phi^{\text{in}}=\hat{I}\,f(\epsilon)$. By the same token, 
Eqs.~\eqref{eq:totalHeat} and \eqref{eq:ExpansionPhi} give
\begin{equation}
I_{\text{tot}}^{Q}=\int_{-\infty}^{\infty}\frac{d\epsilon}{2\pi}(\epsilon-\mu)\,\tr_c\!\left\{ \phi^{\text{out}(1)}+\phi^{\text{out}(2)}\right\} \,.
\label{eq:totalHeatSimplified}
\end{equation}
These expressions nicely elucidate the connection between heat and entropy 
currents, and the departure from $dQ=Td\boldsymbol{S}$ beyond the quasistatic limit. 
At first order in $\dot X$, corresponding to the quasistatic regime, the entropy 
current is entirely given by the heat current over temperature $I_{\text{tot}}^{S(1)}=I_{\text{tot}}^{Q(1)}/T$, i.e., the proposed form of the entropy current correctly
connects to the quasistatic equilibrium. In contrast, at second order
an additional negative correction appears
\begin{align}
I_{\text{tot}}^{S(2)} & =\frac{I_{\text{tot}}^{Q(2)}}{T}+
\int_{-\infty}^{\infty}\frac{d\epsilon}{2\pi}\frac{1}{2T\partial_{\epsilon}f}\tr_c\left\{ \left(\phi^{\text{out}(1)}\right)^{2}\right\} \,.\label{eq:IS2}
\end{align}
Since $\tr_c\{(\phi^{\text{out}(1)})^{2}\}$ contains all off-diagonal elements of 
$\phi^{\text{out}(1)}$, it encodes the correlations created by the dynamic scatterer. 
These correlations determine by how much the entropy current in the leads 
is smaller than the corresponding heat current over temperature. This net inflow
of entropy into the scattering region reflects the local dissipation-induced
increase of entropy.

We calculate $\phi$ explicitly within the gradient expansion \cite{BodePRL,Bode2012,Thomas2012}.
Assuming that $f_{\alpha}(\epsilon)=f(\epsilon)$,
one writes $\phi^{\text{out}(1)}$ in terms of the frozen scattering matrix
$S$,
\begin{align}
\phi^{\text{out}(1)}\left(\epsilon,t\right) & =i\dot{X}\partial_{\epsilon}f\,S\partial_{X}S^{\dagger}\,.
\label{eq:Phi1Main}
\end{align}
Inserting $\phi^{\text{out}(1)}$ into the entropy current Eq.\ \eqref{eq:IS2}, we obtain the entropy current up to second order
\begin{equation}
I_{\text{tot}}^{S}=\frac{I_{\text{tot}}^{Q}}{T}-\frac{\dot{W}^{(2)}}{T}\,.\label{eq:EntropyTotWdiss}
\end{equation}
with
\begin{equation}
\dot{W}^{(2)}=-\frac{\dot{X}^{2}}{2}\,\int_{-\infty}^{\infty}\frac{d\epsilon}{2\pi}\partial_{\epsilon}f(\epsilon)
\tr_c\left(\partial_{X}S^{\dagger}\partial_{X}S\right)\,\geq0\,.
\label{eq:Dissipation}
\end{equation}
Remarkably, $\dot{W}^{(2)}=\gamma\dot{X}^{2}$ is exactly the dissipated power that the external agent pumps into the system as a result of the time-dependent system Hamiltonian. 
$\dot{W}^{(2)}$ was derived in Refs.~\cite{BodePRL,Bode2012,Thomas2012} in terms of the friction coefficient  $\gamma$  of
the back-action force that needs to be overcome by the external agent. 
Thus, from our \emph{outside perspective,} dissipation leads to an
inflow of entropy into the scattering region in addition to the
heat-current contribution. 

We are now ready to discuss the inside-outside duality of entropy evolution: We utilize the acquired knowledge about the entropy current (outside perspective) to draw conclusions about the evolution of the entropy $\boldsymbol{s}$ of the strongly coupled subsystem located in the scattering region (inside perspective). The direct calculation of the thermodynamic functions of such a subsystem has proven problematic in the past due to difficulties in taking proper account of the coupling Hamiltonian and the presence of strong hybridization \cite{LilianaDynamical,EspositoQThermo,Bruch2016,Ochoa2016}. These problems are naturally avoided within the Landauer-B\"uttiker formalism.
Since this formalism considers fully coherent unitary
dynamics in both the leads and the scattering region, the von-Neumann
entropy associated with the scattering states is conserved in a scattering
event. Hence, an additional inflow of entropy is reflected in an
increased entropy $\boldsymbol{s}$ stored in the scattering region. As a result, the 
entropy is source-free 
\begin{equation}
\frac{d\boldsymbol{s}}{dt}+I_{\text{tot}}^{S}=0\,.\label{eq:SourceFreeS}
\end{equation}
We can use this continuity equation and Eq.\ \eqref{eq:EntropyTotWdiss} to infer the evolution of $\boldsymbol{s}$.
Invoking energy and particle conservation, we identify $\dot{Q}=-I_{\text{tot}}^{Q}$
as the heat leaving the scattering region from the inside perspective.
Thus, the entropy evolution can be expressed in terms of the thermodynamic functions 
of the (strongly) coupled subsystem as
\begin{equation}
\frac{d\boldsymbol{s}}{dt}=\frac{\dot{Q}}{T}+\frac{\dot{W}^{(2)}}{T}\,.\label{eq:SecondLaw}
\end{equation}
Therefore, dissipation leads to a local increase of entropy, which is
provided by the scattered electrons. This constitutes the inside-outside
duality of entropy evolution. 

Integrated over a full cyclic transformation of $X$, the entropy
current needs to vanish, as it derives from a source-free thermodynamic state function, see
Eq.\ \eqref{eq:SourceFreeS}. Averaged over a cycle,
Eq.\ \eqref{eq:EntropyTotWdiss} thus implies that all extra
energy pumped into the scattering region $\dot{W}^{(2)}$ eventually has
to be released as heat into the leads
\begin{equation}
\overline{I_{\text{tot}}^{Q(2)}}=\overline{\dot{W}^{(2)}}\,.
\end{equation}


{\it Application to the resonant level model.$-$}To emphasize the advantage of the outside approach over calculating the thermodynamic functions of a subsystem directly, we connect here to the thermodynamics of the resonant level model derived earlier from an \emph{inside perspective}.
This model consists of a single localized electronic
level $H_{D}=\varepsilon_{d}(t)d^{\dagger}d$, which can be changed
in time by an external agent. It is coupled to a free electron metal
$H_{B}=\sum_{k}\epsilon_{k}c_{k}^{\dagger}c_{k}$ via a coupling Hamiltonian
$H_{V}=\sum_{k}\left(V_{k}d^{\dagger}c_{k}+\text{h.c.}\right)$ and
was intensively studied in the past \cite{LilianaDynamical,Ochoa2016},
with difficulties in Ref.\ \cite{EspositoQThermo} pointed out and
overcome in Ref.\ \cite{Bruch2016}. 

The inside approach demands a splitting of the coupling Hamiltonian $H_{V}$
between effective system and bath, which strongly limits its applicability to the resonant level model in the wide band limit of energy-independent hybridization \cite{Bruch2016,Ochoa2016}.
In contrast, the here developed outside approach yields the strong coupling thermodynamics for arbitrary non-interacting electron systems and furthermore reproduces the results for the resonant level.
Deriving the distribution
matrix $\phi$ for this model explicitly, we show order by order that both the heat current $I^{Q}$ in Eq.\ \eqref{eq:totalHeatSimplified} and  the entropy current $I_{\text{tot}}^{S}$ in Eq.\ \eqref{eq:IStot} exactly reproduce the absorbed heat $\dot{Q}=-I_{\text{tot}}^{Q}$ and entropy change $\dot{\boldsymbol{s}}=-I_{\text{tot}}^{S}$ from the inside perspective \cite{Bruch2016} (see Supplemental Material).
Thereby we also explicitly confirm the inside-outside duality of entropy evolution: The dissipated power $\dot{W}^{(2)}$ was shown to lead to a local
increase of entropy for the resonant level in Ref.\ \cite{Bruch2016}, and we demonstrate here that this is reflected in an additional inflow of entropy $I_{\text{tot}}^{S}$ carried by the scattering states, leaving the entropy source-free, Eq.\ \eqref{eq:SourceFreeS}.



{\it Conclusion.$-$}We developed a Landauer-B\"uttiker approach to entropy evolution in
strongly coupled fermionic systems, which considers fully coherent
quantum dynamics in combination with coupling to macroscopic equilibrium
baths. This formalism naturally avoids the system-bath distinction
and is applicable to arbitrary non-interacting electron systems. We
showed that the entropy current generated by a dynamic scatterer depends
on the correlations between different scattering channels, which are
generated in the scattering event. At quasistatic order, the entropy
current is just the heat current over temperature, while at next order
the dissipation induced by the finite velocity transformation yields
a net inflow of entropy into the scattering region. This inflow reflects
the dissipation-induced local increase of entropy constituting the inside-outside
duality of entropy evolution.


{\it Acknowledgments.$-$}We acknowledge fruitful discussions with A. Nitzan. This work was
supported by the Deutsche Forschungsgemeinschaft (SFB 658), 
the German Academic Exchange Service (DAAD), and the Conselho 
Nacional de Pesquisa e Desenvolvimento (CNPq).


\newpage 
\appendix
\onecolumngrid

\section{Supplemental Material}

\title{Supplemental Material for  
\\ \textit{ Landauer-B\"uttiker approach to strongly coupled quantum thermodynamics: inside-outside duality of entropy evolution}}

\author{Anton Bruch}
\affiliation{\mbox{Dahlem Center for Complex Quantum Systems and Fachbereich Physik, Freie Universit\"at Berlin, 14195 Berlin, Germany}}
\author{Caio Lewenkopf}
\affiliation{Instituto de F\'{i}sica, Universidade Federal Fluminense, 24210-346 Niter\'{o}i, Brazil}
\author{Felix von Oppen}
\affiliation{\mbox{Dahlem Center for Complex Quantum Systems and Fachbereich Physik, Freie Universit\"at Berlin, 14195 Berlin, Germany}}

\date{\today}

\maketitle

\section{Calculation of the outgoing distribution matrix in the gradient expansion
\label{sec:TimeDepDistributionFct}}

In the following we derive the adiabatic expansion for the outgoing
distribution matrix Eq.\ (4) of the main text.
With the expression of the outgoing operators $b$ in terms of the
incoming ones $a$ via exact scattering matrix of the time-dependent
problem ${\cal S}$, Eq.\ (2) of the main text, we obtain
\begin{align*}
\braket{b_{\beta}^{\dagger}\left(\epsilon_{2}\right)b_{\alpha}\left(\epsilon_{1}\right)} & =\sum_{\gamma\delta}\int\frac{d\epsilon_{3}}{2\pi}\int\frac{d\epsilon_{4}}{2\pi}\big<\,{\cal S}_{\beta\gamma}^{*}(\epsilon_{2},\epsilon_{3})a_{\gamma}^{\dagger}(\epsilon_{3}){\cal S}_{\alpha\delta}(\epsilon_{1},\epsilon_{4})a_{\delta}(\epsilon_{4})\,\big>\,.
\end{align*}
We use that the incoming scattering states are uncorrelated equilibrium
channels
\begin{equation}
\braket{a_{i}^{\dagger}(\epsilon_{1})a_{j}(\epsilon_{2})}=\delta_{ij}2\pi\delta(\epsilon_{1}-\epsilon_{2})f_{i}(\epsilon_{1})
\end{equation}
and get 
\begin{align}
\braket{b_{\beta}^{\dagger}\left(\epsilon_{2}\right)b_{\alpha}\left(\epsilon_{1}\right)} & =\sum_{\gamma}\int\frac{d\epsilon_{3}}{2\pi}\int\frac{d\epsilon_{4}}{2\pi}\bigg\{{\cal S}_{\alpha\gamma}(\epsilon_{1},\epsilon_{3})\tilde{f}_{\gamma}(\epsilon_{3},\epsilon_{4})\,{\cal S}_{\gamma\beta}^{\dagger}(\epsilon_{4},\epsilon_{2})\bigg\}\,,
\end{align}
with $\tilde{f}_{\gamma}(\epsilon_{3},\epsilon_{4})\equiv2\pi\delta(\epsilon_{3}-\epsilon_{4})\,f(\epsilon_{3})$.
The Wigner transform of a convolution
\begin{equation}
G\left(\epsilon_{1},\epsilon_{2}\right)=\int\frac{d\epsilon_{3}}{2\pi}C(\epsilon_{1},\epsilon_{3})D(\epsilon_{3},\epsilon_{2})
\end{equation}
takes the form of a Moyal product of Wigner transforms 
\begin{equation}
G\left(\epsilon,t\right)=C\left(\varepsilon,t\right)*D\left(\varepsilon,t\right)
\end{equation}
where $C\left(\varepsilon,t\right)*D\left(\varepsilon,t\right)=C\left(\varepsilon,t\right)\exp\left[\frac{i}{2}\left(\stackrel{\leftarrow}{\partial}_{\varepsilon}\vec{\partial}_{t}-\stackrel{\leftarrow}{\partial}_{t}\vec{\partial}_{\varepsilon}\right)\right]D\left(\varepsilon,t\right)$.
Hence, we get

\begin{align}
\phi_{\alpha\beta}^{\text{out}}(\epsilon,t)=&\int\frac{d\tilde{\epsilon}}{2\pi}e^{-i\tilde{\varepsilon}t} \left \langle b_{\beta}^{\dagger}(\epsilon-\tilde{\epsilon}/2)b_{\alpha}(\epsilon+\tilde{\epsilon}/2) \right\rangle \\
=&\sum_{\gamma}\left[\mathfrak{{\cal S}}_{\alpha\gamma}(\epsilon,t)*\tilde{f}_{\gamma}(\epsilon,t)\right]*\mathfrak{{\cal S}}_{\gamma\beta}^{\dagger}(\epsilon,t)\,.\label{eq:PhiOutWigner}
\end{align}
Expanding the exponential gives the different orders of velocity,
which is called the gradient expansion. The Wigner transform of the incoming
distribution in channel $\gamma$ $\tilde{f}_{\gamma}(\epsilon_{3},\epsilon_{4})=2\pi\delta(\epsilon_{3}-\epsilon_{4})\,f_{\gamma}(\epsilon_{3})$
is just the Fermi function of the associated reservoir $\tilde{f}_{\gamma}(\epsilon,t)=f_{\gamma}(\epsilon)$.
The Wigner transform of the full scattering matrix 
\begin{equation}
\mathfrak{{\cal S}}(\epsilon,t)=\int\frac{d\tilde{\epsilon}}{2\pi}\e^{-i\tilde{\epsilon}t}\mathfrak{{\cal S}}(\epsilon+\frac{\tilde{\epsilon}}{2},\epsilon-\frac{\tilde{\epsilon}}{2})
\end{equation}
can be written as an expansion in powers of velocity (assuming $\ddot{X}=0$)\cite{Moskalets2004AdQPump,BodePRL} 
\begin{equation}
\mathfrak{{\cal S}}(\epsilon,t)=S_{t}(\epsilon)+\dot{X}A_{t}(\epsilon)+\dot{X}^{2}B_{t}(\epsilon)\,,\label{eq:AdiabaticExpSMatrix}
\end{equation}
where $S$ is the frozen scattering matrix, $A$ is the A-matrix,
its first order correction, and $B$ is its second order correction.
All these matrices depend parametrically on time and from now on we
drop their energy and time labels for better readability. The second
order contribution to the scattering matrix $B$ never contributes
to the distribution matrix up to second order in $\dot{X}$ in absence
of a bias $f_{\alpha}=f\forall\alpha$, as we show below. 

In the setting of a quantum dot $H_{D}=\sum_{n,n'}d_{n}^{\dagger}h_{n,n'}(X)d_{n'}$
 coupled to leads $H_{L}=\sum_{\eta}\epsilon_{\eta}c_{\eta}^{\dagger}c_{\eta}$
via a coupling Hamiltonian $H_{T}=\sum_{\eta,n}c_{\eta}^{\dagger}W_{\eta n}d_{n}+h.c.$, the frozen scattering matrix $S$ can be expressed in terms of the
frozen retarded Green's function of the of the quantum dot $G^{R}$
and the coupling matrices $W$ between the dot and the attached leads
\begin{equation}
S=1-2\pi i\nu WG^{R}W^{\dagger}\,,\label{eq:MahauxWeidenmueller}
\end{equation}
where $\nu$ is the density of states in the leads. This formula is
called the Mahaux-Weidenmueller formula. 

In the case of an energy-independent hybridization the $A$-matrix
can be written as \cite{BodePRL}
\begin{equation}
A=\pi\nu W\left(\partial_{\epsilon}G^{R}\Lambda_{X}G^{R}-G^{R}\Lambda_{X}\partial_{\epsilon}G^{R}\right)W^{\dagger}\,,\label{eq:A-Matrix}
\end{equation}
where $\Lambda_{X}=\partial h(X)/\partial X$.

\subsection{Zeroth order}

At zeroth order only the frozen scattering matrix in the zeroth order
gradient expansion contributes

\begin{align}
\phi_{\alpha\beta}^{\text{out}(0)} & =\sum_{\gamma}S_{\alpha \gamma}f_{\gamma}S_{\gamma \beta}^{\dagger}\,.
\end{align}
In the absence of voltage and temperature bias $f_{\alpha}^{\text{in}}=f\forall\alpha$,
this simplifies to
\begin{align}
\phi_{\alpha\beta}^{\text{out}(0)} & =f\sum_{\gamma}S_{\alpha \gamma}S_{\gamma \beta}^{\dagger}=f\,\left[SS^{\dagger}\right]_{\alpha\beta}\,.
\end{align}
For $f_{\alpha}^{\text{in}}=f\forall\alpha$, we can order by order simplify the expression for the outgoing distribution matrix $\phi^{\rm out}_{\alpha\beta}$ by invoking the unitarity of the scattering matrix at the corresponding order. At
zeroth order this condition is just the unitarity of the frozen scattering
matrix
\begin{equation}
SS^{\dagger}=\hat{I}\,,
\end{equation}
which leads to the zeroth order outgoing distribution matrix diagonal
in channel lead space, identical to the incoming distribution matrix
\begin{equation}
\phi_{\alpha\beta}^{\text{out}(0)}=\delta_{\alpha\beta}f=\phi_{\alpha\beta}^{\text{in}}\,.\label{eq:phi_0}
\end{equation}
We derive the unitarity conditions for the first and second order
below in Sec. \ref{subsec:Unitarity-condition-at}.

\subsection{First order}

At first order there are contributions both from the zeroth order
gradient expansion with the first order correction to $\mathfrak{\mathfrak{{\cal S}}}(\epsilon,t)$,
$\dot{X}A_{t}^{X}$ in Eq.\ \eqref{eq:AdiabaticExpSMatrix}, and the
first order gradient expansion with the frozen scattering matrix $S$, which is simplified by the
fact that the incoming distribution function $f_{\gamma}$ has no time
dependency. We obtain

\begin{align}
\phi_{\alpha\beta}^{\text{out}(1)} & =\dot{X}\sum_{\gamma}\bigg[A_{\alpha \gamma}f_{\gamma}S_{\gamma \beta}^{\dagger}+S_{\alpha \gamma}f_{\gamma}A_{\gamma \beta}^{\dagger}\nonumber \\
 & +\frac{i}{2}\left\{ \partial_{\epsilon}S_{\alpha \gamma}\partial_{X}S_{\gamma \beta}^{\dagger}-\partial_{X}S_{\alpha \gamma}\partial_{\epsilon}S_{\gamma \beta}^{\dagger}\right\} f_{\gamma}\nonumber \\
 & +\frac{i}{2}\left\{ -\partial_{X}S_{\alpha \gamma}\partial_{\epsilon}f_{\gamma}S_{\gamma \beta}^{\dagger}+S_{\alpha \gamma}\partial_{\epsilon}f_{\gamma}\partial_{X}S_{\gamma \beta}^{\dagger}\right\} \,\bigg]\,,
\end{align}
where we used $\partial_{t}S=\dot{X}\partial_{X}S$.

If we assume $f_{\alpha}=f\forall\alpha$ we get

\begin{align}
\frac{\phi_{\alpha\beta}^{\text{out}(1)}}{\dot{X}} & =f\sum_{\gamma}\left[A_{\alpha \gamma}S_{\gamma \beta}^{\dagger}+S_{\alpha \gamma}A_{\gamma \beta}^{\dagger}+\frac{i}{2}\left\{ \partial_{\epsilon}S_{\alpha \gamma}\partial_{X}S_{\gamma \beta}^{\dagger}-\partial_{X}S_{\alpha \gamma}\partial_{\epsilon}S_{\gamma \beta}^{\dagger}\right\} \right]\nonumber \\
 & +\partial_{\epsilon}f\sum_{\gamma}\frac{i}{2}\left\{ -\partial_{X}S_{\alpha \gamma}S_{\gamma \beta}^{\dagger}+S_{\alpha \gamma}\partial_{X}S_{\gamma \beta}^{\dagger}\right\} \,.
\end{align}
This can be simplified by the unitarity condition at first order Eq.\ 
\eqref{eq:unitarity conditionFirstOrder}

\begin{align}
\phi_{\alpha\beta}^{\text{out}(1)} & =\dot{X}\partial_{\epsilon}f\sum_{\beta=LR}\frac{i}{2}\left\{ -\partial_{X}S_{\alpha \gamma}S_{\gamma \beta}^{\dagger}+S_{\alpha \gamma}\partial_{X}S_{\gamma \beta}^{\dagger}\right\} \nonumber \\
 & =\dot{X}\partial_{\epsilon}f\,i\left[S\partial_{X}S^{\dagger}\right]_{\alpha\beta}\label{eq:phi_1}
\end{align}
where we used $\partial_{X}\left(SS^{\dagger}\right)=\partial_{X}\hat{I}=0$
.

\subsection{Second order }

At second order there are contributions by the zeroth order gradient
expansion with $\dot{X}^{2}B$ in Eq.\ \eqref{eq:AdiabaticExpSMatrix},
the first order gradient expansion with $\dot{X}A$ and the second
order gradient expansion with $S$. Using Eq.\ \eqref{eq:unitaritySecondOrder}
we can simplify the expression to 

\begin{align}
\frac{\phi_{\alpha\beta}^{\text{out}(2)}}{\dot{X}^{2}} & =\frac{1}{2}\partial_{\epsilon}^{2}f\left[\partial_{X}S\partial_{X}S^{\dagger}\right]_{\alpha\beta}\nonumber \\
 & +\partial_{\epsilon}f\frac{i}{2}\left[A\partial_{X}S^{\dagger}+S\partial_{X}A^{\dagger}+\frac{i}{2}\left(\partial_{X}^{2}S\partial_{\epsilon}S^{\dagger}+\partial_{\epsilon}S\partial_{X}^{2}S^{\dagger}-\partial_{\epsilon}\partial_{X}S\partial_{X}S^{\dagger}-\partial_{X}S\partial_{X}\partial_{\epsilon}S^{\dagger}\right)\right]_{\alpha\beta}\,.\label{eq:phi2}
\end{align}

\subsection{Unitarity condition at different orders
\label{subsec:Unitarity-condition-at}}

The unitarity of the full scattering matrix
\begin{equation}
\sum_{n}\int\frac{d\epsilon}{2\pi}\mathfrak{{\cal S}}_{mn}(\epsilon',\epsilon)\mathfrak{{\cal S}}_{nk}^{\dagger}(\epsilon,\epsilon'')=2\pi\delta(\epsilon'-\epsilon'')\delta_{mk}
\end{equation}
leads to different conditions at each order in velocity. Taking the
Wigner transform of this expression leads to
\begin{equation}
1\delta_{mk}=\sum_{n}\mathfrak{{\cal S}}_{mn}(\epsilon,t)*\mathfrak{{\cal S}}_{nk}^{\dagger}(\epsilon,t)\,.\label{eq:UnitarityFullS}
\end{equation}
We insert the adiabatic expansion of the scattering matrix Eq.\ \eqref{eq:AdiabaticExpSMatrix}
and consistently collect the terms order by order in the velocity.

\paragraph{Zeroth order}

To zeroth order in the velocity we obtain the unitarity condition
for the frozen scattering matrix
\begin{equation}
\delta_{mk}=\sum_{n}S_{mn}S_{nk}^{\dagger}\,.\label{eq:unitarityZerothOrder}
\end{equation}

\paragraph{First order}

Up to first order in the velocity Eq.\ \eqref{eq:UnitarityFullS} reads
\begin{align}
\delta_{mk} & =\sum_{n}\bigg[S_{mn}S_{nk}^{\dagger}+\dot{X}A_{mn}S_{nk}^{\dagger}+\dot{X}S_{mn}A_{nk}^{,\dagger}(\epsilon)\nonumber \\
 & +\frac{i}{2}\left(\frac{\partial S_{mn}}{\partial\epsilon}\frac{\partial S_{nk}^{\dagger}}{\partial t}-\frac{\partial S_{mn}}{\partial t}\frac{\partial S_{nk}^{\dagger}}{\partial\epsilon}\right)\,\bigg]\,.
\end{align}
Using that the frozen scattering matrix $S$ is unitary and $\partial_{t}S=\dot{X}\partial_{X}S$
this yields
\begin{equation}
\sum_{n}\left[A_{mn}^{X}S_{nk}^{\dagger}+S_{mn}(\epsilon)A_{nk}^{X,\dagger}\right]=-\sum_{n}\frac{i}{2}\left(\frac{\partial S_{mn}}{\partial\epsilon}\frac{\partial S_{nk}^{\dagger}}{\partial X}-\frac{\partial S_{mn}}{\partial X}\frac{\partial S_{nk}^{\dagger}}{\partial\epsilon}\right)\,.\label{eq:unitarity conditionFirstOrder}
\end{equation}

\paragraph{Second order}

Up to second order in the velocity one obtains upon using Eqs. \eqref{eq:unitarityZerothOrder}
and \eqref{eq:unitarity conditionFirstOrder} 

\begin{align}
0 & =\dot{X}^{2}\sum_{n}\bigg[A_{mn}A_{nk}^{\dagger}+S_{mn}B_{nk}^{,\dagger}(\epsilon)+B_{mn}S_{nk}^{\dagger}\nonumber \\
 & +\frac{i}{2}\left(\frac{\partial A_{mn}}{\partial\epsilon}\frac{\partial S_{nk}^{\dagger}}{\partial X}-\frac{\partial A_{mn}}{\partial X}\frac{\partial S_{nk}^{\dagger}}{\partial\epsilon}\right)+\frac{i}{2}\left(\frac{\partial S_{mn}}{\partial\epsilon}\frac{\partial A_{nk}^{\dagger}}{\partial X}-\frac{\partial S_{mn}}{\partial X}\frac{\partial A_{nk}^{\dagger}}{\partial\epsilon}\right)\nonumber \\
 & -\frac{1}{8}\left(\partial_{\epsilon}^{2}S_{mn}\partial_{X}^{2}S_{nk}^{\dagger}+\partial_{X}^{2}S_{mn}\partial_{\epsilon}^{2}S_{nk}^{\dagger}-2\partial_{\epsilon}\partial_{X}S_{mn}\,\partial_{X}\partial_{\epsilon}S_{nk}^{\dagger}\right)\bigg]\,.\label{eq:unitaritySecondOrder}
\end{align}

\section{Application to the resonant level model\label{AppApplicResLevel}}

Here we derive the distribution matrix $\phi$ for the resonant level
model introduced in the main text. In this
case the scattering matrix can be reduced to a single element, the reflection
coefficient, which can be obtained via the Mahaux-Weidenmueller formula
Eq.\ \eqref{eq:MahauxWeidenmueller}
\begin{align}
S(\epsilon,t) & =1-\frac{i\Gamma}{\epsilon-\epsilon_{d}(X)+i\Gamma/2}\,.\label{eq:SResonantLevel}
\end{align}
Here $\Gamma$ is the decay rate of the dot electrons into the lead
$\Gamma=2\pi\sum_{k}\left|V_{k}\right|^{2}\delta\left(\varepsilon-\varepsilon_{k}\right)$,
and the $A$-matrix Eq.\ \eqref{eq:A-Matrix} vanishes \cite{Bode2012}.
The distribution matrix $\phi$ only contains a single element
describing the occupation of the in- and outgoing scattering channel.
The first order contribution $\phi^{\text{out}(1)}$ Eq.\ \eqref{eq:phi_1}
takes the form 

\begin{align}
\phi^{\text{out}(1)}\left(\epsilon,t\right) & =-\dot{X}\partial_{\epsilon}f\frac{\partial\epsilon_{d}}{\partial X}A_{dd}\,,\label{eq:Phi1ResLevel}
\end{align}
while the second order Eq.\ \eqref{eq:phi2} can be simplified to 
\begin{equation}
\phi^{\text{out}(2)}\left(\epsilon,t\right)=\dot{X}^{2}\frac{1}{2}\partial_{\epsilon}^{2}f\left(\frac{\partial\epsilon_{d}}{\partial X}\right)^{2}A_{dd}^{2}\,,\label{eq:Phi2ResLevel}
\end{equation}
where $A_{dd}=\Gamma/\left([\epsilon-\epsilon_{d}(X)]^{2}+\Gamma^{2}/4\right)$
is the spectral function of the dot electrons.

With this we show that the heat current in the leads $I^{Q}$
Eq.\ (6) of the main text is identical to the heat that leaves
the extended level from the inside perspective $\dot{Q}$ order by
order \cite{Bruch2016}

\begin{align}
I^{Q(1)} & =-\dot{\epsilon}_{d}\int_{-\infty}^{\infty}\frac{d\epsilon}{2\pi}\,(\epsilon-\mu)\partial_{\epsilon}fA_{dd}=-\dot{Q}^{(1)},\\
I^{Q(2)} & =\frac{1}{2}\dot{\varepsilon}_{d}^{2}\int_{-\infty}^{\infty}\frac{d\epsilon}{2\pi}\,(\epsilon-\mu)\partial_{\epsilon}^{2}fA_{dd}^{2}=-\dot{Q}^{(2)}\,,
\end{align}
where we wrote $\dot{X}\frac{\partial\epsilon_{d}}{\partial X}=\dot{\epsilon}_{d}$
to directly compare to the quantities in Ref.\ \cite{Bruch2016}.

The inside-outside duality of entropy evolution Eq.\ (24) of the main text
can be explicitly checked order by order by inserting $\phi^{\text{out}}$
Eqs. \eqref{eq:Phi1ResLevel} and \eqref{eq:Phi2ResLevel} into $I_{\text{tot}}^{S}$
Eq.\ (18) of the main text and compare it to the change of the entropy $\boldsymbol{s}$ 
of the resonant level in Ref.\ \cite{Bruch2016}
\begin{align}
I^{S(1)} & =-\dot{\epsilon}_{d}\int_{-\infty}^{\infty}\frac{d\epsilon}{2\pi}\,\frac{(\epsilon-\mu)}{T}\partial_{\epsilon}fA_{dd}=-\frac{d\boldsymbol{s}}{dt}^{(1)}\\
I^{S(2)} & =\dot{\varepsilon}_{d}^{2}\int_{-\infty}^{\infty}\frac{d\epsilon}{2\pi}\,\left\{ \frac{(\epsilon-\mu)}{T}\frac{1}{2}\partial_{\epsilon}^{2}fA_{dd}^{2}+\frac{\partial_{\epsilon}f}{2T}A_{dd}^{2}\right\} =-\frac{d\boldsymbol{s}}{dt}^{(2)}\,.
\end{align}
Thereby we reproduced the results of Ref.\ \cite{Bruch2016} with the
here developed outside perspective.

\section{Entropy current and entanglement\label{sec:Entanglement}}

The developed form of the entropy current carried by the scattering
states allows us to quantify the the correlations between different
scattering channels created in the scattering event. If we regard
the case of two attached leads ($L$ and $R$), we can measure the
correlations created by the scattering event in terms of the quantum
mutual information between outgoing channels to the left and right 
\begin{equation}\label{MutualInfo}
{\cal I}(L:R)=\boldsymbol{S}_{L}+\boldsymbol{S}_{R}-\boldsymbol{S}_{\text{tot}}\,,
\end{equation}
where $\boldsymbol{S}{}_{L/R}$ is the von-Neumann entropy of the
reduced density matrix of the left (right) lead and $\boldsymbol{S}_{\text{tot}}$
is the total entropy of the outgoing states, including correlations
between $L$ and $R$. In the case of a pure state of the composite
system $S_{\text{tot}}=0$, and ${\cal I}(L:R)$ reduces to twice the entanglement
entropy ${\cal F}=S_{L}=S_{R}$ created in the scattering event. 

The reduced density matrix and the corresponding entropy of the outgoing
states in the left and right lead can be obtained by a method developed
by Peschel \cite{Peschel2003,Peschel2012}, which takes the form of
the argument presented in the main text,
but confined to the subspace of interest. This results in the entropy
of the reduced density matrix of subsystem $A$ 
\begin{equation}
\boldsymbol{S}_{A}=\text{tr}_c\left(\sigma[\phi^{A}]\right)\,.
\end{equation}
where $\phi^{A}$ is the submatrix of the full distribution matrix
$\phi$, defined on subspace $A$ only. Applied to outgoing scattering
states analogous to the derivation in the main text, we obtain from Eq.\ \eqref{MutualInfo} a
mutual information current 

\begin{align}
I^{MI} & =I_{\text{red},L}^{S}+I_{\text{red},R}^{S}-I_{\text{tot}}^{S,\text{out}}\,,\label{eq:MutualInfoCurrent}
\end{align}
as a measure of the correlations created per unit time. Here
\begin{equation}
I_{\text{tot}}^{S,\text{out}}=\int_{-\infty}^{\infty}\frac{d\epsilon}{2\pi}\,\tr_c\left\{ \sigma[\phi^{\text{out}}(t,\epsilon)]\right\} \,,
\end{equation}
is the outgoing component of the total entropy current Eq.\ (15) of the main text.
The entropy current corresponding to the reduced density matrix of
the electrons in the left (right) lead $I_{\text{red},L(R)}^{S}$ takes the
form
\begin{equation}
I_{\text{red},L(R)}^{S}=\int_{-\infty}^{\infty}\frac{d\epsilon}{2\pi}\,\tr_c\left\{ \sigma[\phi^{\text{out},L(R)}(t,\epsilon)]\right\} \,,
\end{equation}
with $\phi^{\text{out},L(R)}$ being the submatrix of the distribution matrix
$\phi^{\text{out}}$ defined on the left (right) subspace only.

The simplest case to which we can apply this developed formalism is
the case of a static scatterer between two reservoirs at zero temperature
with an applied bias voltage, which was investigated as a device to
create entangled electron-hole pairs in Ref.\ \cite{Beenakker2003}.
Considering two channels on each side (the authors of Ref.\ \cite{Beenakker2003}
consider a quantum Hall setup, in which the two channels can either
represent two spin channels within the same Landau level or two different
Landau levels), the scattering matrix,
\begin{equation}
S=\left(\begin{array}{cc}
r & t'\\
t & r'
\end{array}\right)\,,
\end{equation}
is a $4\times4$ matrix, with the $2\times2$ submatrices $r$,$r'$,
$t$, $t'$ describing the reflection and transmission from the left
or right respectively. Here we neglect the energy dependence of the
scattering matrix in the bias window and drop the energy labels for
better readability.

For a static scatterer, the outgoing distribution matrix can be obtained
by a simplified version of Eq.\ \eqref{eq:PhiOutWigner}
\begin{equation}
\phi^{\text{out}}(\epsilon)=S\,\phi^{\text{in}}(\epsilon)\,S^{\dagger}\,.\label{eq:PhiOutStatic}
\end{equation}
Hence in the static case $\phi^{\text{out}}$ is obtained from $\phi^{\text{in}}$
by a unitary transformation given by the frozen scattering matrix
$S$. At zero temperature, the incoming electrons are either fully
occupied or empty at each energy $f_{L}=\Theta(\mu+eV-\epsilon)$
and $f_{R}=\Theta(\mu-\epsilon)$, and hence carry no entropy
\begin{equation}
I_{\text{tot}}^{S\,\text{in}}=\int_{-\infty}^{\infty}\frac{d\epsilon}{2\pi}\,\text{tr}_c\left\{ \sigma[\phi^{\text{in}}(\epsilon)]\right\} =0\,.
\end{equation}
$\phi^{\text{out}}$ in Eq.\ \eqref{eq:PhiOutStatic} then shows
that full outgoing entropy current also vanishes
\begin{equation}
I_{\text{tot}}^{S\,\text{out}}=\int_{-\infty}^{\infty}\frac{d\epsilon}{2\pi}\,\text{tr}_c\left\{ \sigma[\phi^{\text{out}}(\epsilon)]\right\} =0\,,
\end{equation}
since the unitary transformation with the frozen scattering matrix
leaves the incoming pure states pure at each energy. 

For the reduced entropy currents we calculate the outgoing distribution
matrix from Eq.\ \eqref{eq:PhiOutStatic}
\begin{equation}
\phi^{\text{out}}=\Theta(\mu-\epsilon)\hat{I}+\Theta(\mu+eV-\epsilon)\Theta(\epsilon-\mu)\varrho\,,
\end{equation}
with
\begin{equation}
\varrho=\left(\begin{array}{cc}
rr^{\dagger} & rt^{\dagger}\\
tr^{\dagger} & tt^{\dagger}
\end{array}\right)\,.
\end{equation}
Inserting $\phi^{\text{out}}$ into the mutual information current Eq.\ \eqref{eq:MutualInfoCurrent}
leads to

\begin{align}
I^{MI} & =\frac{eV}{2\pi}\tr_c\sigma\left[rr^{\dagger}\right]+\frac{eV}{2\pi}\tr_c\sigma\left[tt^{\dagger}\right]\\
 & =2\frac{eV}{2\pi}\tr_c\sigma\left[tt^{\dagger}\right]\\
 & =2\frac{eV}{2\pi}\left(\sigma(T_{1})+\sigma(T_{2})\right)\,,
\end{align}
where $T_{1},T_{2}\in\left(0,1\right)$ are the eigenvalues of the
transmission matrix product $t^{\dagger}t=\hat{I}-r^{\dagger}r$,
which reproduces the entanglement entropy or entanglement of formation
${\cal F}=I^{MI}/2$ calculated in Ref.\ \cite{Beenakker2003}.

\end{document}